\documentclass[letterpaper]{article}
\usepackage{aaai} 
\usepackage{times} 
\usepackage{helvet} 
\usepackage{courier}
\usepackage{graphicx,paralist}
\usepackage{amssymb}
\usepackage{amsmath}
\usepackage{adjustbox}

\nocopyright

\title{A General Framework for Anytime Approximation in Probabilistic Databases}
\author{Maarten Van den Heuvel \and Floris Geerts\\
University of Antwerp, Belgium\\
\AND
Wolfgang Gatterbauer\\
Northeastern University, Boston, MA, USA\\
\And
Martin Theobald\\
University of Luxembourg, Luxembourg\\}
\newcommand{\cref}[1]{\ref{#1}}
\usepackage{balance}  
\usepackage[usenames,table,dvipsnames,x11names]{xcolor}      

\def\eox{\unskip\kern 10pt{\unitlength1pt\linethickness{.4pt}$\diamondsuit${}}}

\newcommand{\squishlist}{
 \begin{list}{$\bullet$}
  { \setlength{\itemsep}{0pt}
     \setlength{\parsep}{5pt}
     \setlength{\topsep}{5pt}
     \setlength{\partopsep}{0pt}
     \setlength{\leftmargin}{1.5em}
     \setlength{\labelwidth}{1em}
     \setlength{\labelsep}{0.5em} } }
 \newcommand{\squishend}{\end{list}}

\usepackage{balance}
\usepackage{mathtools}	
\usepackage{graphicx}     	
\usepackage{multirow}       
\usepackage{amsmath}				
\usepackage{amssymb}                
\usepackage{bm}			
\usepackage{upgreek}       

\usepackage{caption}                
\usepackage{xspace}            
\usepackage[caption=false]{subfig}   

\usepackage{enumitem}               

\usepackage{color}	
\usepackage{colortbl}             

\usepackage{tabularx}		

\usepackage{rotating}

\usepackage{aliascnt}

\usepackage{tikz}
\usepackage{tikz-qtree}			

\usetikzlibrary{positioning,chains,fit,shapes,calc}

\let\footnotesize\scriptsize

\newaliascnt{lemma}{theorem}				
              	
\aliascntresetthe{lemma}  					

\newaliascnt{conjecture}{theorem}			
    
\aliascntresetthe{conjecture}  				

\newaliascnt{remark}{theorem}				
              
\aliascntresetthe{remark}  					
\newaliascnt{corollary}{theorem}			
      
\aliascntresetthe{corollary}  				

\newaliascnt{definition}{theorem}			
    
\aliascntresetthe{definition}  				

\newaliascnt{proposition}{theorem}			
  
\aliascntresetthe{proposition}  				
\newaliascnt{example}{theorem}			
  	
\aliascntresetthe{example}  				
\newaliascnt{problem}{theorem}

\aliascntresetthe{problem}

\usepackage[usenames,table,dvipsnames]{xcolor}      

\newcounter{result}

\newsavebox{\coloredbgbox}

\newcommand{\specificref}[2]{\hyperref[#2]{#1~\ref*{#2}}}			
\newcommand{\specialref}[3]{\hyperref[#2]{#1~\ref*{#2}~#3}}

\newcommand{\figref}[2]{\hyperref[#1]{Fig.~\ref*{#1}#2}}	
\newcommand{\Figref}[2]{\hyperref[#1]{Figure~\ref*{#1}#2}}	

\newcommand{\figuresref}[2]{\hyperref[#1]{Fig.~\ref*{#1} to \ref*{#2}}}	
\newcommand{\Figuresref}[2]{\hyperref[#1]{Figures~\ref*{#1} to \ref*{#2}}}

\usepackage[algo2e,ruled, vlined]{algorithm2e}		
\DontPrintSemicolon     		
\SetNlSty{}{}{}					
\SetAlgoCaptionSeparator{\,}	
\SetAlgoSkip{smallskip}

\SetAlgoCaptionLayout{myCap}					

\SetAlCapNameSty{textbf}
\SetAlCapNameFnt{\small}
\SetAlgoCaptionSeparator{:}

\SetInd{1.2mm}{1.2mm}	

\SetKwFor{ForAll}{forall}{do}{endfch}	
\LinesNumbered

\newcommand{\PPP}{{{\mathbb{P}}}}

\renewcommand{\epsilon}{\varepsilon}

\newcommand{\smallsection}[1]{\vspace{3mm}\noindent\textbf{#1.}}	
\newcommand{\introparagraph}[1]{\textbf{#1.}}        

\definecolor{gray}{rgb}{0.5,0.5,0.5}
\definecolor{niceblue}{rgb}{.8,.85,1}

\newcommand{\futurework}[1]{}

\newcommand{\datarule}{{\,:\!\!-\,}}

\newcommand{\makeop}[2]                         
  {\ifx#2.\def\next##1{}\else\escapechar=-1     
  \def\next##1{\escapechar=92\def#2{#1}}        
  \expandafter\next\expandafter{\string#2}      
  \let\next\makeop\fi\next{#1}}                 

\makeop{{\cal #1}}                              
  \W  .             

\def \var(#1){{\bf #1}}

\newcommand{\silentreminder}[1]{}

\newcommand{\eat}[1]{}

\graphicspath{{figs/}}

\newif\ifqed



\begin{document}

\maketitle

\begin{abstract}
\begin{quote}
Anytime approximation algorithms that compute the probabilities of queries over probabilistic databases can be of great use to statistical learning tasks. 
Those approaches have been based so far on either ($i$) sampling 
or ($ii$) branch-and-bound with model-based bounds.
We present here a more general branch-and-bound framework 
that extends the possible bounds by using ``dissociation,'' which yields tighter bounds.
\end{quote}
\end{abstract}

\section{Introduction}

Since calculating the exact probability of a query over probabilistic databases (PDBs) is \#P-hard in general, 
probabilistic inference is a major bottleneck in several related applications, such as statistical relational learning \cite{RaedtK17}, 
and fast approximation methods are needed.
\emph{Anytime approximation methods} 
give approximate answers fast, yet
allow the user to refine the answer by using additional time and resources.
The current state of the art in anytime approximation for PDBs are either based on ($i$) sampling \cite{re2007efficient}
or on ($ii$) the model-based branch-and-bound approach by \cite{DBLP:journals/vldb/FinkHO13} implemented in the SPROUT system, where the latter outperforms the former. 
See \cite{DBS-052} for a recent survey.

We propose here to improve the branch-and-bound approach by replacing model-based bounds with novel \emph{dissociation-based bounds}, which were shown to dominate model-based bounds~\cite{gatterbauer2014oblivious}.
The technique of dissociation is related to the variable splitting framework by \cite{choi2010relax},
and has so far only been applied at the first-order (query) level for upper bounds~\cite{DBLP:journals/vldb/GatterbauerS17}.
One reason is that for conjunctive queries, these dissociation-based upper bounds are uniquely defined and proven to be better than any model-based bounds. 
In contrast, a whole spectrum of optimal oblivious lower bounds exists, which includes the model-based bounds as special cases. 
Important ingredients of our approach are strategies for quickly choosing good lower bounds, as well as a novel heuristic for the ``branch'' part of the algorithm based on \emph{influence} or sensitivity \cite{kanagal2011sensitivity}.

\section{Background}

\introparagraph{The problem}
We consider the evaluation of Boolean conjunctive queries (CQs)
and illustrate our problem and approach with the following query $Q$ over PDB $D$:
\begin{align*}
	Q \datarule R(X), S(X,Y), T(Y)
\end{align*}
\begin{center}

\setlength{\tabcolsep}{1mm}
\begin{tabular}{c}
\end{tabular}
$D$
\hspace{3mm}
\begin{tabular}{|c|c |c|}
\multicolumn{1}{c}{$R$} & \multicolumn{1}{c}{$X$} & \multicolumn{1}{c}{$p$}\\
\hline
$r_1$ & a & $0.5$\\
$r_2$ & b & $0.6$\\
\hline
\multicolumn{3}{c}{}
\end{tabular}
\hspace{1mm}
\begin{tabular}{|c|c c|c|}
\multicolumn{1}{c}{$S$} & \multicolumn{1}{c}{$X$} & \multicolumn{1}{c}{$Y$} & \multicolumn{1}{c}{$p$}\\
\hline
$s_1$ & a & c & 0.3\\
$s_2$ & a & d & 0.4\\
$s_3$ & b & d & 0.5\\
\hline
\end{tabular}
\hspace{1mm}\begin{tabular}{|c|c|c|}
\multicolumn{1}{c}{$T$} & \multicolumn{1}{c}{$Y$} & \multicolumn{1}{c}{$p$}\\
\hline
$t_1$ & c & 0.4\\
$t_2$ & d & 0.8\\
\hline
\multicolumn{3}{c}{}
\end{tabular}
\end{center}

\noindent
Grounding $Q$ over $D$ leads to a propositional formula $\varphi$ that is called the ``lineage of $Q$ over $D$'' in which each variable represents a tuple in the database. In our example, 
$
\varphi=r_1(s_1t_1 \vee s_2t_2) \vee r_2s_3t_2
$. 
Computing its probability $\PPP(\varphi)$ is equivalent to computing the probability $\PPP(Q)$, i.e.\ the probability that the query $Q$ is true over PDB $D$. 
In general, calculating this probability is \#P-hard in the number of variables in the lineage, and thus in the number of tuples in the database.
We are interested in developing an approximation scheme that allows us to trade-off available time with required accuracy of approximation.

\medskip
\introparagraph{State of the art}
The algorithm underlying SPROUT \cite{DBLP:journals/vldb/FinkHO13}  approximates
$\PPP(\varphi)$ by using lineage decompositions based on independence and determinism. 
When the smaller lineages obtained are read-once (i.e.\ have no multiple occurrences of the same variable) their probabilities are computed exactly in PTIME, otherwise they are approximated by \emph{model-based bounds}: 
for each variable, all except one of its occurrences are set to true or 1 (resp.\ false or 0) while the remaining occurrence is assigned the original probability to get an upper (resp.\ lower) bound. 
SPROUT randomly selects these occurrences.
The bounds are propagated back up, based on the decomposition, to  obtain an approximation of $\PPP(\varphi)$.
If this approximation is not accurate enough, Shannon expansion (SE) is applied on the decomposed lineages: a variable $x$ is selected and, based on 
$
\varphi \equiv (x \wedge \varphi|_{x=1}) \vee (\neg x \wedge \varphi|_{x=0})
$, the decomposition and approximation process continues on $\varphi|_{x=1}$ and $\varphi|_{x=0}$
until the desired accuracy is obtained. SPROUT selects the most frequent variable (the one with the highest number of occurrences, ties broken randomly) for SE. Clearly, the challenge is to limit the number of SEs, as these can double the size of the formulas, resulting in higher computation cost.

\section{Our approach}
The branch-and-bound approach just described is generic: any technique for approximating the probability of lineages can be plugged in during decomposition, and similarly, any variable selection procedure for SE can be used. 

We propose to use \emph{dissociation} to obtain approximations.
Dissociation replaces multiple occurrences of a variable with independent copies. 
Our example lineage $\varphi$ becomes $\varphi'=r_1(s_1t_1 \vee s_2t_2') \vee r_2s_3t_2''$. 
By carefully assigning new probabilities $p_U$ and $p_L$ to both copies $t_2'$ and $t_2''$ of $t_2$, $\PPP_{p_U}(\varphi')$ is guaranteed to be an upper, 
and $\PPP_{p_L}(\varphi')$ to be a lower bound for $\PPP(\varphi)$. 
What makes dissociations particularly attractive for our framework is that dissociation bounds generalize model-based bounds, and that they can be calculated efficiently.
As a consequence, we obtain better approximations in each step, hereby possibly reducing the number of SEs needed to obtain a desired accuracy.
For SE, we propose a method based on influence. We next detail some aspects of our approach and highlight some of its advantages.

\begin{figure}[t]
\vspace{-3ex}
\subfloat[]{
	\includegraphics[scale=0.36]{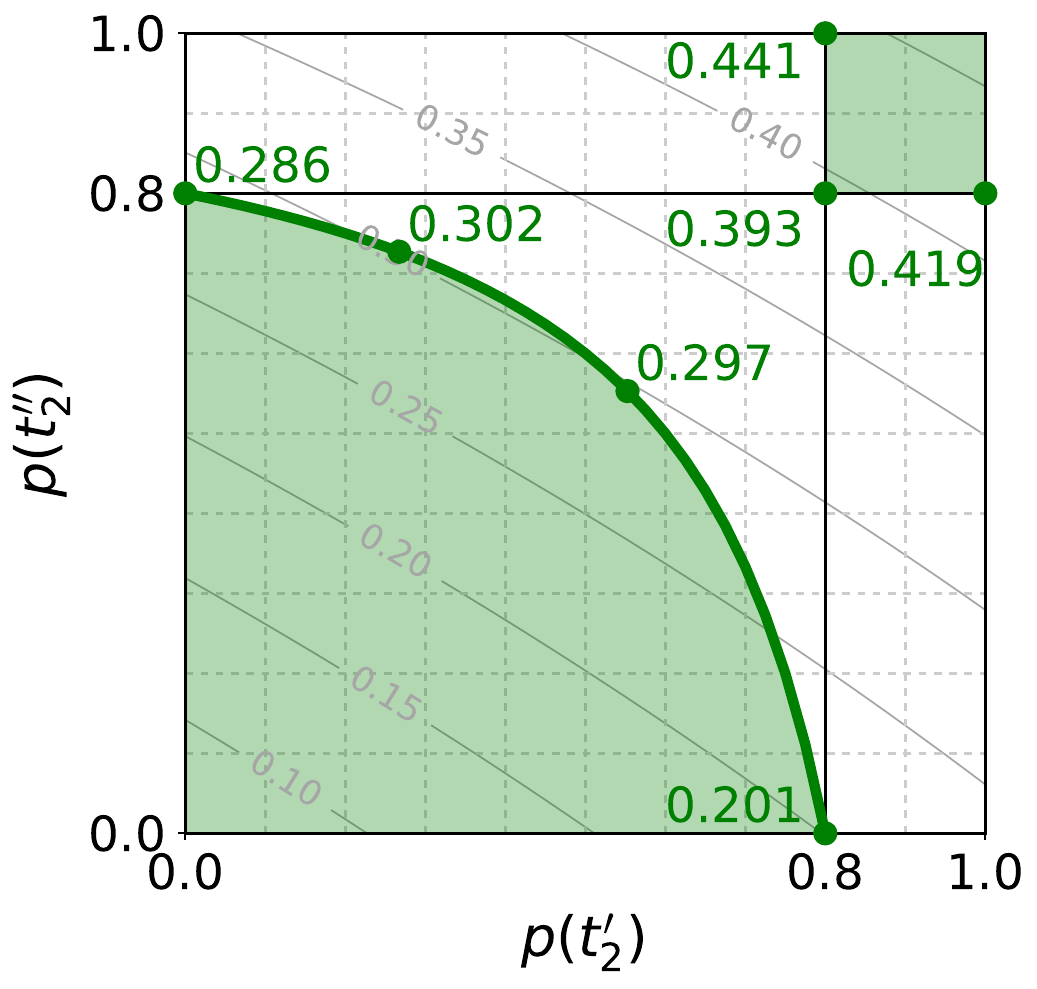}
	\label{fig:example}}
\subfloat[]{
	\includegraphics[scale=0.26]{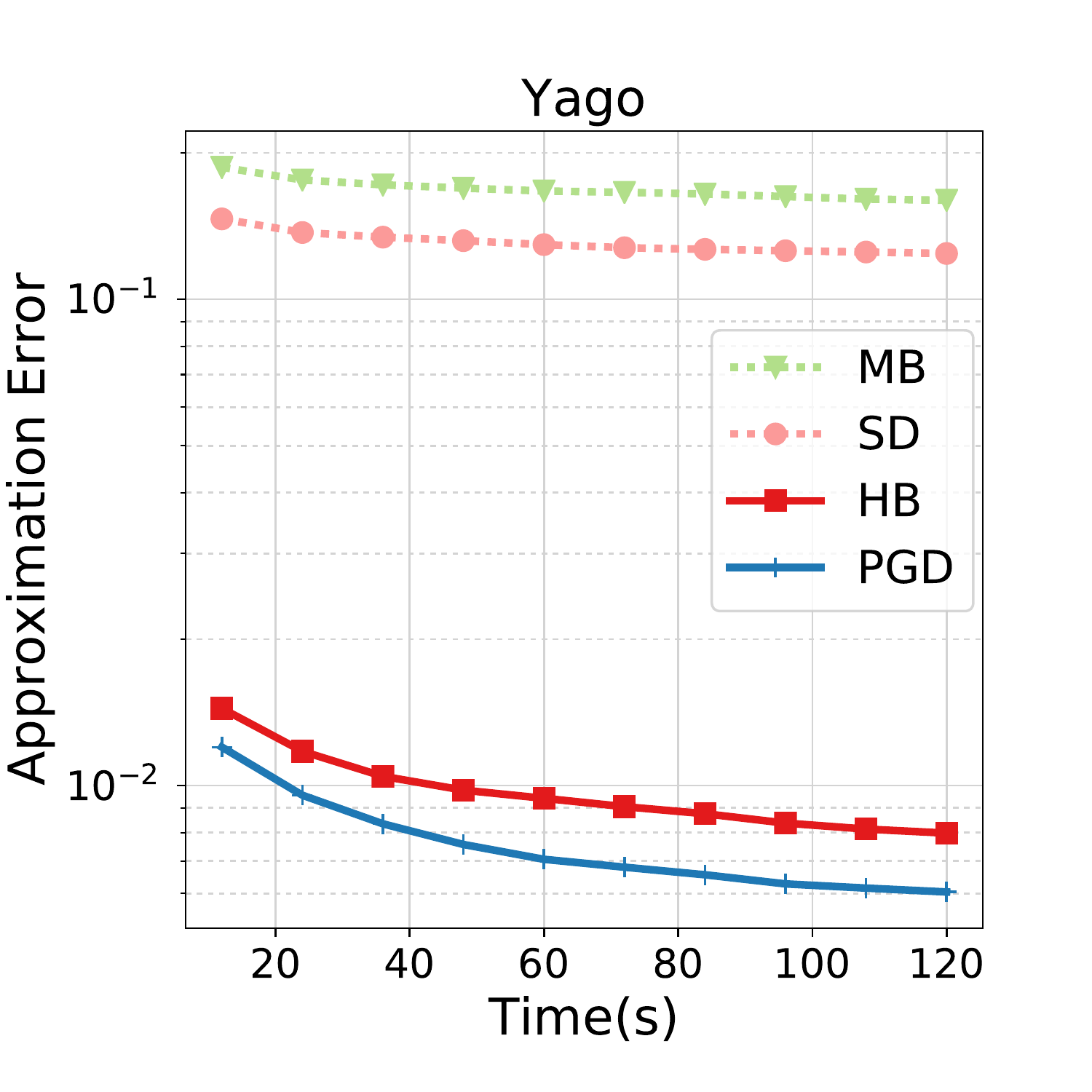}
	\label{fig:experiment}}
	\vspace{-1ex}
\caption{a) Green shaded areas show all possible assignments for upper and lower dissociation bounds for our running example. 
Green values show resulting approximations.
b) Our new methods (PBD and HB) achieve better accuracy-time tradeoffs than prior methods (MB and SD) on the Yago3 dataset.}
\vspace{-3ex}
\end{figure}

\introparagraph{1. Better bounds}
Following Theorem 4.8 in ~\cite{gatterbauer2014oblivious},
the two green shaded areas in Fig.\ \ref{fig:example} show all possible assignments of probabilities to the occurrences $t_2'$ and $t_2''$ in our lineage $\varphi'$
that guarantee to upper or lower bound the true probability $\PPP(\varphi)=0.384$.
For upper bounds, the two model-based bounds provide an approximation
of $0.419$ and $0.441$. By contrast, assigning the original probability $p(t_2)=0.8$ to both $t_2'$ and $t_2''$ results in the 
unique and optimal upper dissociation bound of $0.393$.

For the lower bounds, the shaded area on the lower left shows all possible lower bound assignments, and all assignments on the curved border are possible optima.
Here, the two model-based bounds are just a few of the possible options, achieving approximations of $0.201$ and $0.286$ respectively. 
The \emph{symmetric} lower dissociation bound assigns both occurrences an equal share by setting $p_L(t_2') = p_L(t_2'') = 1 - \sqrt[2]{1-p(t_2)}$ which gives $0.297$ as lower bound. 
The best assignment lies slightly to the left and gives $0.302$.
 
Part of our approach are \emph{gradient-descent methods} that aim to find this optimal assignment. 
Computation of the gradient
is closely related to the notion of \emph{influence},
which can be computed in PTIME for read-once formulas~\cite{kanagal2011sensitivity}, as well as for dissociated formulas. 
While gradient-descent methods generally take longer to find the best lower bound than randomly assigning model-based bounds, the quality of the bound is often much better, thus justifying these additional computations.

\introparagraph{2. Better variable selection}
The state-of-the-art heuristic for choosing a variable for SE in SPROUT is selecting the \emph{most frequent} variable. 
We instead select the variable for SE that has the highest \emph{sum of influences} of each of its occurrences, as if they were independent. 
This ensures PTIME computation and turns out to be a better choice. 
Moreover, we can re-use the computed influences from our gradient-based optimization methods, thus avoiding re-computation.

\introparagraph{3. Optimization trade-off}
We have a  ``knob'' to control
how much time to spend on finding a good lower bound using gradient descent, before moving on to a next Shannon expansion.  
Recall that model-based bounds are fast to compute; 
they are randomly selected, in one step. 
In contrast, descent methods perform multiple steps,
and this extra computation pays off only when considerably better lower bounds are obtained,
and further SEs can be avoided.

\section{Experiments}
We experimented with several approximation
methods on the Yago3 dataset~\cite{DBLP:conf/cidr/MahdisoltaniBS15}.
We obtained 380 lineages from 4 different queries by assigning different input probabilities to the data, using different join-orders to factorize the lineages, and by injecting different constants into the queries. Figure \ref{fig:experiment} shows the average approximation error over time for four different instantiations of our approach. 

\verb+MB+ is the existing Model-based approach from SPROUT and performs slightly worse than \verb+SD+, 
which uses Symmetric Dissociations for both upper and lower bounds.
Both methods work best with the frequency heuristic for SE.
Our new methods \verb+PGD+ (Projected Gradient Descent) and \verb+HB+ (Hybrid method) vastly outperform both approaches. 
Both methods use the optimal symmetric upper bounds and apply 10 gradient descent steps before using SE.
But whereas \verb+PGD+ searches for the true best lower bounds, 
\verb+HB+ searches for the best possible model-based lower bound, by moving in a gradient direction.
This makes the optimization faster, but produces slightly worse bounds than \verb+PGD+.
These methods work best with the influence heuristic for SE, but outperform the others even with the frequency heuristic.

\section{Conclusion}

We introduced an anytime approximation framework for probabilistic query evaluation.
Our framework leverages novel dissociation bounds that generalize and improve upon model-based bounds.
Our experimental results show notable improvements over the current state of the art,
and we believe that the approach also has applications beyond PDBs in the broader area of statistical-relational learning (SRL).

\smallsection{Acknowledgements}

This work has been supported in part by NSF IIS-1762268 and FWO G042815N.
\bibliography{ms} 

\bibliographystyle{aaai}
\end{document}